\documentclass[aps,pre,twocolumn,groupedaddress,showpacs]{revtex4}
\usepackage{graphicx}
\usepackage{dcolumn}
\usepackage{bm}
\usepackage{epsfig}
\input epsf
\epsfclipon

\begin{document}
\title{Scaling of human behavior during portal browsing}
\author{ Anna Chmiel, Kamila Kowalska, and  Janusz A. Ho{\l}yst } \affiliation{Faculty of Physics, Center of Excellence
for Complex Systems Research, Warsaw University of Technology,
Koszykowa 75, PL-00-662 Warsaw, Poland}
\date{\today}
\begin{abstract}

We investigate transitions of portals users between different subpages. A weighted network of portals subpages is reconstructed where edge weights are numbers of corresponding transitions. Distributions of link weights and node strengths follow power laws over several decades. Node strength increases faster than linearly with  node degree. The distribution of time spent by the user at one subpage decays as power law with exponent around 1.3. Distribution of numbers $P(z)$ of unique subpages during one visit   is exponential. We find a square root dependence between the  average $z$ and  the total number of transitions $n$ during a single visit. Individual  path of portal user resembles of self-attracting walk on the weighted network. Analytical model is developed to recover in part the collected data.

\end{abstract}  \maketitle

\section{Introduction}

There are three categories of Internet users: the  observers limit they activity only to be in touch with family and friends, consumers treat the Internet as a source of information which they need  in order to  work and to plan the free time, hobbies, travel etc. The last category are the creators, they are active on newsgroups, forums or  write their own blogs. The distributions of  users  for these three categories change really fast  showing the evolution of the society. In Poland  53 percent of men and 46 percent of women use Internet  while most of them are observers and consumers.

The web portals gained special attention because they are, for many users, the starting point of their web browser. The simple structure with homepage aggregating almost all links  allows  inexperienced  users to find the basic information. For the more practiced user the web portal is a usual habit, every day they follow  news at portals. The set of web pages they choose, the browsing art and the time they spend reading the site's content reflect their pastime habits and interests in the real world. They combined two spaces real and virtual in an equilibrium.  The analysis of the statistics of users behavior can bring a deeper insight  about human dynamics on the web. This issue was thoroughly discussed in articles \cite{pB,Huberman,Gon,ben,natureB}.
Our goal in this paper was to analyze the behavior of users browsing portals (portal consists of subpages and one of them is a home page). Visitors behavior is described by the way they  navigate between subpages, how much time they spend at each of them, if and how many times they come back to the previously visited subpages. Knowledge about these habits can be useful for portals designers, as well as for planing marketing strategy that should fix an optimal number and distribution of advertisements.

We organized the present paper as follows. In Sec .\ref{II} we explain  details of the dataset we use, in Sec. III we present various statistics of portal browsing. We construct a weighted network of subpages and analyze  standard properties  \cite{przeg} of this network. In Sec. IV we focus on the users browsing  strategy. The way of navigating between portal subpages is modeled as a special self-attracting walk and the simulation results are given in Sec.V.
We also develop a simple analytic approach (Sec. VI) that in part fits to numerical data. 

\section{DATASET}\label{II}
 The analysis was based on cookie statistics provided by Gemius company for two Polish portals. Cookie is a small text file that every  web browser automatically receives from a web server while visiting a web-page. It allows to differentiate users and to maintain data related to them during navigation. The shortcoming of this method is that some  users can delete their cookies and get a new one next time they enter the Internet. Therefore the number of cookie users does not correspond to the number of real Internet users. The difference is the larger the longer the time of data collection, being negligible for a day , but immensely significant for a month. For this reason we used for our analysis daily data only. 
Our data was collected between 27/07/07 and 28/07/07 at two  portals with  large numbers of daily visitors (portal A had more than million and portal B around 3.9 million cookie users during twenty-four hours). The internal structure of the portal, i.e its division to subpages, was taken into account. On the day of investigations the portal A consisted of $N=195$ active subpages, where as an active page we counted a subpage that was visited at least once during the analyzed day.  The portal B consisted of $N=515$ active subpages. A number of active pages can vary from day to day.  
The data gathered for one web-site included: user ID, subpage ID, time of user's arrival at a subpage (and more precisely, the time she or he clicked at this page). It should be emphasized that the whole collection of times and subpages ID for a given user, which we will refer to as a "visit chain" in due course, cannot be identified with the time physically spent by the user at the subpages. This is a natural consequence of the fact that we detect users  activity only  by their transition to new   subpages and we do not have any information about their real behavior between two such events. In particular, we are unable to verify whether a user spent the time between two transitions reading the subpage content or whether he left it very quickly,  started other activity (e.g. a phone call) and  then came back to browse the Internet again. 

\section{ NETWORK STRUCTURE AND TRAFFIC AT THE PORTAL }

We  constructed a weighted network of subpages, defining a link weight as a number of users moving from one subpage (vertex) to another. We observed that the number of transitions from a node {\em i} to a node {\em j}, $m(i \rightarrow j)$, is roughly equal to the number of transitions in other directions $m(i \rightarrow j) \simeq m(j \rightarrow i)$. We simplified the network topology by   introducing  undirected links with  weights: 
\begin{equation}\label{w}
w_{ij}= m(i \rightarrow j)+m(j\rightarrow i).
\end{equation}

\begin{figure}
\vskip 0.3cm
 \centerline{\epsfig{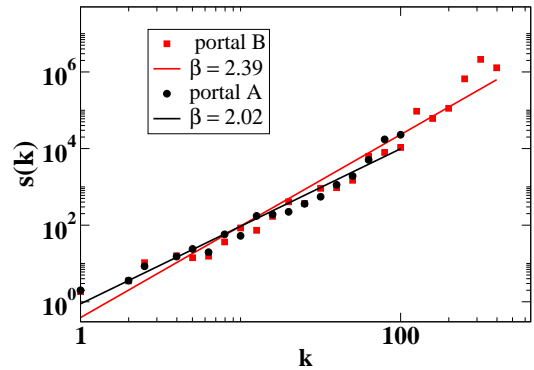}}
    \caption{Relation between average strength of a node and the degree of a node (black circles - portal A; color (red) square - portal B).}
        \label{sk}
\end{figure}

One can observe a frequent  habit to retract to the subpage that was previously visited, so one visitor  can pass many times over the same link. Therefore, the maximum link weight can be larger than the total number of users visiting the portal in the considered time period.
\begin{figure}[ht]
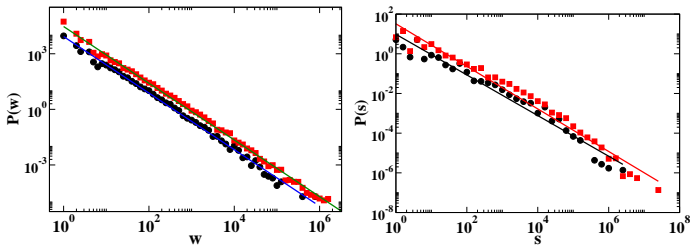

\centerline{\begin{tabular}{ c c}
 \epsfig{file=wagi.eps,width=.25\textwidth} & \epsfig{file=ps.eps,width=.25\textwidth} \\
\end{tabular}}
    \caption{The weight distribution on the right, and the strength distribution on the left  (black circles - portal A; color (red) square - portal B).}
    \label{wag_s}
\end{figure}
The weight distributions are power law with the characteristic exponents $\gamma=1.5$ for both portals. Defining a node  strength  in a usual way:
\begin{equation}\label{w}
s_i= \sum_{j}w_{ij}
\end{equation}
we found that the strength distribution is close to $P(s)\sim 1/s$ in both cases. 
In order to get more precise analysis of the topological properties of the weighted network we should look for the correlations between a node strength $s_i$ and its degree  $k_i$. If a linear relation $s=\langle w \rangle k $ were observed, the correlations would be absent. In our system we find a positive correlation: Fig. \ref{sk} presents a power law dependence $s\sim k^{\beta}$ with the  exponent $\beta$ larger than two thus strengths of vertices grow faster than their degrees. A similar positive correlation was  observed in many real-world networks, e.g.  in the world-wide airport networks \cite{PNAS}. In the work \cite{chin} an explanation of this positive correlation for  a large class of networks is explained as result  of preferential flow allocation. 

\begin{figure}
 \centerline{\epsfig{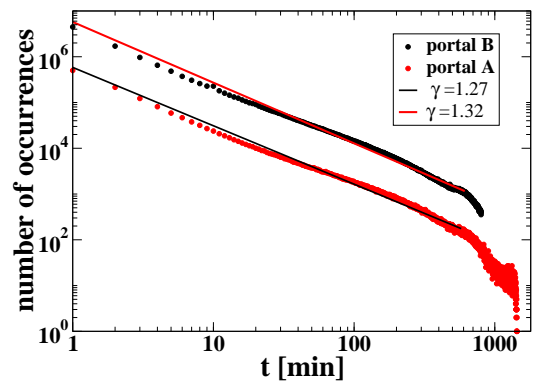}}
    \caption{The distribution of time spent by the user at one subpage.}
        \label{fig_3}
\end{figure}

The time spent by a user at one subpage is understood as the time between two consecutive clicks on two different subpages performed by this user. Corresponding  time distributions were analyzed by Dezs\"{o} et al. \cite{pB} and by Gon\c{c}alves and Ramasco \cite{Gon} who found power law relations with exponents $\gamma=1.2$ and $\gamma=1.25$ respectively. In our case we measured  $\gamma=1.27$ for portal A and $\gamma=1.32$ for portal B and the scaling was valid for the range over two decades (see Fig. \ref{fig_3}).The model of separately  executed tasks  \cite{natureB} gives   $\gamma=1$ while the  model of (bounded) tasks groups  \cite{Gon} leads to $\gamma > 1$.  

\begin{figure}
 \centerline{\epsfig{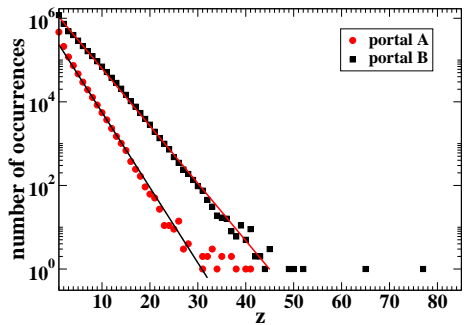}}
    \caption{Distribution of  numbers of unique subpages {\em z} visited by a user during one visit. The parameter $\alpha$  of exponential fitting is 0.31 for portal A and 0.41 for portal B.}
        \label{unikat}
\end{figure}

\begin{figure}
\vskip 0.3cm
 \centerline{\epsfig{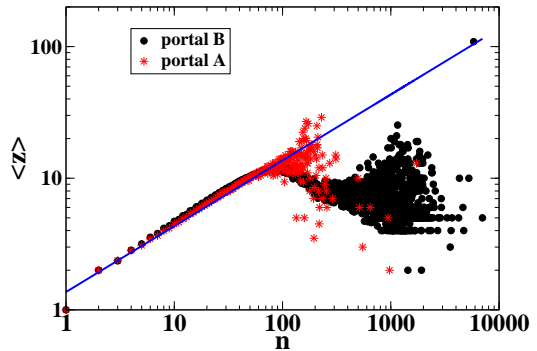}}
    \caption{(color on line) Relation between the average number of distinct subpages  $\langle z\rangle$ and number of jumps $n$. Red point data from portal A, black point data from portal B, line fitted with  $\langle z\rangle=1.44\sqrt{n}$.}
        \label{sqr}
\end{figure}

\begin{figure}
 \centerline{\epsfig{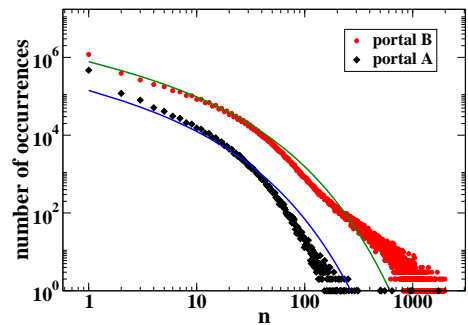}}
    \caption{Jump distribution and fit to Eq.(\ref{pn}).}
   \label{all}
\end{figure}

The weighted network we investigated was created by a group of users. To understand  properties of this network we followed  details of users paths. We analyzed  the distribution of numbers of unique subpages $z$ visited by a user during a single visit (Fig. \ref{unikat}) and we observed an exponential behavior with a unique characteristic parameter $\alpha$ for a given portal: 
\begin{equation}\label{un}
P(z)=A \, \exp(-\alpha \, z)
\end{equation} 
Let us consider the relation between two variables:  a number of jumps (transitions)  between subpages  {\em n}  and  the average number of distinct (unique) subpages {\em $\langle z \rangle$}  corresponding to the same  fixed $n$ value. One can refer to {\em  z} as to the "interest horizon" since it describes the user's tendency to stick to limited  subset of  certain subpages. The relation is presented in Fig. \ref{sqr} and for $n \le 100$ can be described by the function:
\begin{equation}\label{ns_n}
\langle z\rangle=a \sqrt{n}.
\end{equation} 
For both portals we observed the same square root dependence  with exactly the same parameter {\em a}. Now, having Eq.(\ref{ns_n}) and the number distribution  of unique subpages, we can find the formula for the  distribution of number of jumps presented in  Fig .\ref{all}.  Since
\begin{equation}\label{law}
P(n)dn= P(z)dz
\end{equation}
we get:
\begin{equation}\label{pn}
P(n)=\frac{a A\,\exp(-\alpha  \,a \sqrt{n})}{2\,\sqrt{n}}.
\end{equation}
As we observed in Fig. \ref{all} this formula fits very well to the collected data.

\section{VISITORS STRATEGY}
The analysis of statistical properties of the visitor behavior reveals an important phenomenon, which are frequent  returns to a subpage previously visited. One can ask how does a subpage at the portal affect the frequency of the return to this subpage. The most special subpage is the homepage, in most cases being the top of the decision tree of a user. We observed that more than 40 percent of users in portal B, and 28.5 percent in portal A were browsing using a special {\it star strategy}: the  visitor  starts at the homepage, then chooses one of other subpages and returns to the homepage in every odd step. At the even steps user reads one of the other subpages, sometimes revisiting some of them. The strategy can be disturbed by users making some slight deviations from the star path, but still they present the same tendency of returning to the previously visited subpage in a small number of steps.

It is clear that if a user chooses a new subpage at every even step, we will not observe the square root relation [Eq. (\ref{ns_n})] but a linear correlation $\langle z\rangle \sim n/2 $. It means that the {\it star strategy} does not explain the scaling observed in the data. Therefore, we assume that the homepage is not the only subpage revisited during one visit chain. Let $p^{*}$  be a return probability  to a subpage visited one step earlier (meaning that a user reads the same subpage at the step numbers {\em n-1} and {\em n+1}). We restricted our calculation to the range of one hundred steps only, where the scaling is observed. In the Tab. \ref{tab1} we present the total return probability $p^{*}$ and its two components that describe the coming back to the homepage ($p_{home}$) and  to a different subpage ($p_{different}$). The total return probability is $p^{*}=p_{home}+p_{different}$. The large value of  $p^{*}$ can be related to the use of the button "back" or opening of a new "window" and jumping between two open browser windows. 

\begin{table}
\setlength{\tabcolsep}{4.5pt}
\begin{center}
\begin{tabular}{lcc}
\hline\hline  & portal A & portal B \\
\hline
probality $p^{*}$ & 0.54  & 0.57\\
probality $p_{home}$ & 0.29 & 0.27\\
probality $p_{different}$ & 0.25  & 0.30\\

\hline\hline
\end{tabular}
\end{center}
\caption{Comparison of return probabilities  data from portal A and B.} \label{tab1}
\end{table}
\section{ SELF-ATTRACTING WALK AS A BROWSING SCENARIO }
The relation between the average number of distinct subpages  $\langle z\rangle $ and the number of jumps $n$ can be interpreted as the relation between the average number of distinct states and the number of steps (understood as time) in  the problem of random walk. This problem was considered by many authors, see e.g.   \cite{book,RW_BL,RW_CN}. In one dimension a random walk is characterized by the square-root relation between the number of distinct states and the number of visited states, $\langle z\rangle \sim \sqrt n$. From the famous theorem of Polya \cite{Polya} it is known that the probability of returning (at any time) to the starting point by a random walker in a {\em d} - dimensional lattice can be less than one only for $d > 2$. In this sense the dimension $d=2$ is critical for this dynamics, $\langle z\rangle \sim n/\log  n$.  For complex networks with  scale-free degree distribution  there  is   $ \langle z\rangle \sim  n$ (see \cite{RW_CN}).

Various models of biased random walks were analyzed, see e.g.  \cite{bias,fronczak}. A special case is a model of self- attracting walk \cite{stanlay,sweel} where a state which was previously visited is preferred in the next time step but  there are different scenarios of attracting relations. Here we adopt such a model for  portal browsing. We took the topology  of the network with weights between two nodes defined by Eq. (\ref{w}). The dynamics on the network is very simple: each walker starts at the homepage and then, with a transition probability $p_{ij}$, moves to one of the neighboring subpages. The probability of transition from  vertex {\em i} to vertex {\em j} is proportional to the weight of this edge 
\begin{equation} p_{ij}=\frac{w_{ij}}{\sum_{k}^{}w_{ik}}=\frac{w_{ij}}{s_i}.
\end{equation}
 After two initial steps  a leaning  towards return to a previously visited page occurs as follows. In the step $n$ a walker returns to  a node visited at $n-2$ step with probability $p^{*}$, and with probability $1-p^{*}$ he chooses a random neighbor, according to transition probability $p_{ij}$. The results of such a simulation for both portals (Fig. \ref{ns12}) are very close to the real data in the range of first thirty steps. 

\section{ANALYTIC APPROXIMATION}
Let us now consider a random walk at the weighted network. In the infinite time limit the stationary occupation probability $\rho_i$ describing the probability that a walker is located at the node {\em i} is given by \cite{stoch_p,InfTeor}:
\begin{equation}\label{sym}
\rho_i=\frac{s_i}{N\langle s\rangle}.
\end{equation}

Unfortunately, in our case the walker life time at the network is not long enough to assume the stationary distribution of probability $\rho_i$  because at the beginning  of the walk the initial site is significant.
\begin{figure*}[ht]
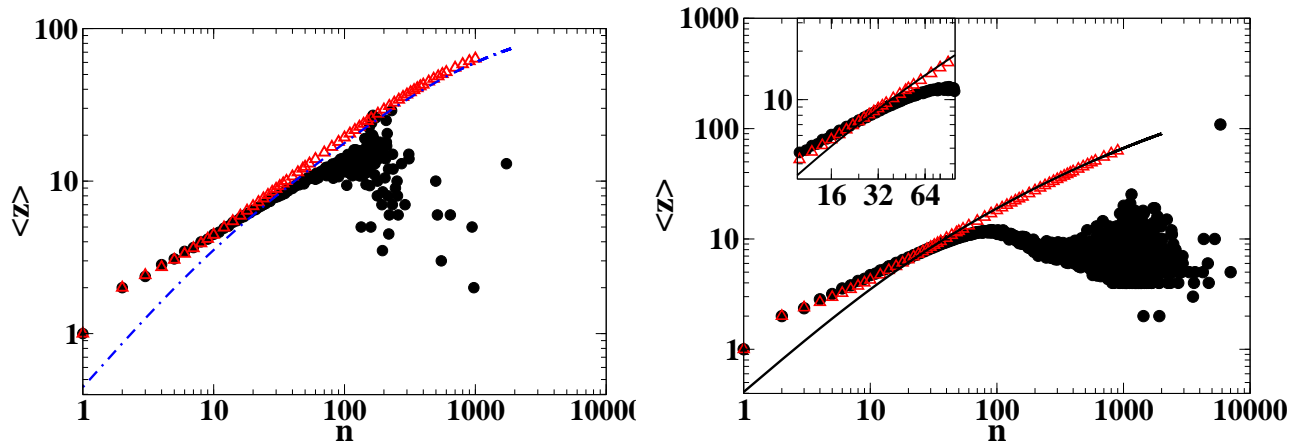

\centerline{\begin{tabular}{ c c}
 \epsfig{file=sat_g_w.eps,width=.47\textwidth} & \epsfig{file=sat1_w.eps,width=.47\textwidth} 
\end{tabular}}
   \caption{ (Color on-line) The relation between the average number of distinct subpages and number of jumps. Portal A on the right  and portal B - on the left; black points are data, lines  are Eq. (\ref{ds2}), red triangles come from the numerical simulation. Results of numerical simulation (1 million  artificial users and  $n=1000$ steps for each) fit perfectly to real data from portal for number of steps $ n<30 $. After system  termalization (more that 30 steps)  we observed the agreement between the analytic calculation and simulations results.}
   \label{ns12}
\end{figure*}
Despite this non-stationarity, we find an approximated  relation between the  average number of distinct subpages $\langle z\rangle $ and the number of jumps  $n$. Let  us  define $d_s(n)$ as a fraction of vertices of  strength $s$ visited by a random walker at least once after $n$ time steps. In our weighted networks the relation between $\langle z\rangle $ and $n$ is analogical to the  unweighted networks discussed   in  \cite{RW_ct}:

\begin{equation}\label{sum}
\langle z\rangle (n)=N\sum_s{P(s)d_s(n)},
\end{equation}
where N is the number of verticies - subpages of the portal. Changes of $d_s(n)$ are given as
\begin{equation}\label{ds}
\frac{\partial{d_s}}{\partial{n}}=[1-d_s(n)]\rho(s).
\end{equation}
Here $\rho(s)$ is the probability that at a walker observed in a random time moment is at site of  strength $s$. The  above equation is true when $\rho(s)$ is stationary  and it is only an approximation  during   first steps of the walker's path. 
Now we extend this equation by taking into account the {\it star strategy} discussed in the previous section.  Since the walker returns to the page visited two steps earlier with probability $p^{*}$  thus in such cases there is  $\frac{\partial{d_s}}{\partial{n}}=0$. It leads to the equation 
\begin{equation}\label{ds1}
 \frac{\partial{d^{*}_s}}{\partial{n}}   =[1-d^{*}_s(n)]\rho(s) (1-p^{*}).
\end{equation}
Taking into account  the initial condition $d^{*}_s(0)=0$ we get the solution 
\begin{equation}\label{ds2}
 d^{*}_s(n) =1-\exp \left[ -n\rho(s)(1-p^{*}) \right]
\end{equation}
with the characteristic  relaxation time 
\begin{equation}\label{tal}
\tau(s) = \frac{1}{\rho(s)(1-p^{*})}.
\end{equation}
One can see that the relaxation process is slowed down by  the factor $1-p^{*}$.
From Eqs. (\ref{sum}) and (\ref{ds2}) we have:
\begin{equation}\label{ds3}
\langle z\rangle(n) = N-N \sum_s  P(s)\exp \left[ -\frac{  s n(1-p^{*})  }{  \langle s\rangle N   }   \right].
\end{equation}
Since for infinite networks there is a divergence of the first moment of the empirically observed distribution $P(s) \sim 1/s$   we used real data to estimate  $\langle s\rangle =\frac{\sum_{i=1}^N{s_i}}{N} $. Resulting solution  (\ref{ds3}) is presented in   Fig. \ref{ns12} and it fits well to numerical simulations discussed in Sec. V.
 \section{CONCLUSIONS}
We show that a one-day user's {\it interest horizon}, measured as the number of distinct visited subpages, is relatively small in comparison to the number of all transitions at the portals, i.e. to the  number of all subpages visited by the user. It means that people return  many times to the same subpage or pass by the same page during one -day visit session. There can be  various explanations of this phenomenon. The large probability of coming back to the homepage can be a result of news portal structure, with a homepage being  a network hub. However, since the probability of returning to any other subpage is also significant, it can suggest that it is somehow difficult for the users to find the information they need. So, if they consider a visited subpage inadequate to what they were expecting, they come back one step up the portal structure and try to go to the other subpage. That could be a hint for the web-designers regarding  the portal functionality. The other conclusion is that the range of the information seeking by  the portals users is in fact restricted only to a narrow range of what they find most interesting. In this sense, Internet seems to be a perfect tool to keep an eye on changing situation in the regions of some importance to the users (for example stock market indexes, political news, topical portals). However, the existence of such a global and easily accessed 'knowledge mine' does not necessarily enlarge people's general interest horizon. The other possibility can result from the fact that portal visitors need frequently to pass over a few "transit" pages to come from one aim to another. Such a scenario would correspond well to the observed value of $\gamma >1$ (see Fig. \ref{fig_3}) of times between consecutive clicks and the model of bounded group tasks proposed in \cite{Gon}.          

Our simple model of a self--attracting walk shows that the real data is very well reproduced  by a short memory process. 
The observed scaling relation between an average number of distinct subpages  $\langle z\rangle$ and a number of jumps $n$  is well reconstructed using the strength of node as a  popularity range and the rule of coming back to the previous page with probability $p^{*}$. The solution of the rate equation fits well to  simulation results for a number of clicks larger than $n>30$. However, it is difficult to directly compare the developed model with the collected  portal data because the number of users visiting more than 30 subpages in not large. This compliance allows us to understand the behavior of the users as a random walk with short memory.

\begin{acknowledgments}
The work was supported by a special grant of Warsaw University of Technology.
\end{acknowledgments}

\end{document}